\documentclass[runningheads]{llncs}

\usepackage{cite}
\usepackage{amsmath,amssymb,amsfonts}
\usepackage{algorithmic}
\usepackage{graphicx}
\usepackage{hyperref}
\usepackage{textcomp}
\usepackage{xcolor}
\usepackage{float}
\usepackage{caption}
\usepackage{subcaption}
\usepackage{lineno}
\usepackage{booktabs}
\usepackage{enumitem}
\usepackage{multirow}
\usepackage{hyperref}

\usepackage{array}
\newcolumntype{L}[1]{>{\raggedright\let\newline\\\arraybackslash\hspace{0pt}}m{#1}}
\newcolumntype{C}[1]{>{\centering\let\newline\\\arraybackslash\hspace{0pt}}m{#1}}
\newcolumntype{R}[1]{>{\raggedleft\let\newline\\\arraybackslash\hspace{0pt}}m{#1}}

\begin{document}

\title{Universal Loss Reweighting to Balance Lesion Size Inequality in 3D Medical Image Segmentation}

\titlerunning{Universal Loss Reweighting to Balance Lesion Size Inequality}

\author{
    Boris Shirokikh \inst{1, 2, 3} \and
    Alexey Shevtsov \inst{2, 3} \and
    Anvar Kurmukov \inst{2, 4} \and
    Alexandra Dalechina \inst{5} \and
    Egor Krivov \inst{2, 3}
    Valery Kostjuchenko \inst{5} \and
    Andrey Golanov \inst{6} \and
    Mikhail Belyaev \inst{1}
}

\authorrunning{B. Shirokikh, A. Shevtsov et al}

\institute{
    Skolkovo Institute of Science and Technology, Moscow, Russia
    \and
    Kharkevich Institute for Information Transmission Problems, Moscow, Russia
    \and
    Moscow Institute of Physics and Technology, Moscow, Russia
    \and
    Higher School of Economics, Moscow, Russia
    \and 
    Moscow Gamma-Knife Center, Moscow, Russia
    \and 
    Burdenko Neurosurgery Institute, Moscow, Russia
    \\
    \email{boris.shirokikh@phystech.edu}
}

\maketitle

\begin{abstract}
    Target imbalance affects the performance of recent deep learning methods in many medical image segmentation tasks. It is a twofold problem: class imbalance – positive class (lesion) size compared to negative class (non-lesion) size; lesion size imbalance – large lesions overshadows small ones (in the case of multiple lesions per image). While the former was addressed in multiple works, the latter lacks investigation. \setcounter{footnote}{0}
We propose a loss reweighting approach to increase the ability of the network to detect small lesions. During the learning process, we assign a weight to every image voxel. The assigned weights are inversely proportional to the lesion volume, thus smaller lesions get larger weights.
    We report the benefit from our method for well-known loss functions, including Dice Loss, Focal Loss, and Asymmetric Similarity Loss. Additionally, we compare our results with other reweighting techniques: Weighted Cross-Entropy and Generalized Dice Loss. Our experiments show that \textit{inverse weighting} considerably increases the detection quality, while preserves the delineation quality on a state-of-the-art level. We publish a complete experimental pipeline\footnote{\url{https://github.com/neuro-ml/inverse_weighting}} for two publicly available datasets of CT images: LiTS and LUNA16. We also show results on a private database of MR images for the task of multiple brain metastases delineation.

\end{abstract}




\section{Introduction}
\label{sec:intro}

In recent years, convolutional neural networks (CNNs) have become the dominant approach to solve medical image segmentation tasks \cite{litjens2017survey}. 
A wide variety of CNN models, training procedures and loss functions were built under the BRATS \cite{brats} and ISLES \cite{maier2017isles} competitions. The most common way to measure the performance of such a new method is to use segmentation voxel-wise metrics, e.g. Dice Score \cite{bakas2018brats}. However, in the case of multiple lesions per image, clinical tasks also require analyzing algorithm in terms of the detection quality. For instance, all tumors, including the smallest ones, should be found and delineated in the brain stereotactic radiosurgery or in the lung cancer screening process. But since the Dice Score is a voxel-wise metric, it does not differentiate between missing several True Positives in a large lesion or in a small one.

\begin{figure}[h]
    \centering
    \includegraphics[width=\linewidth]{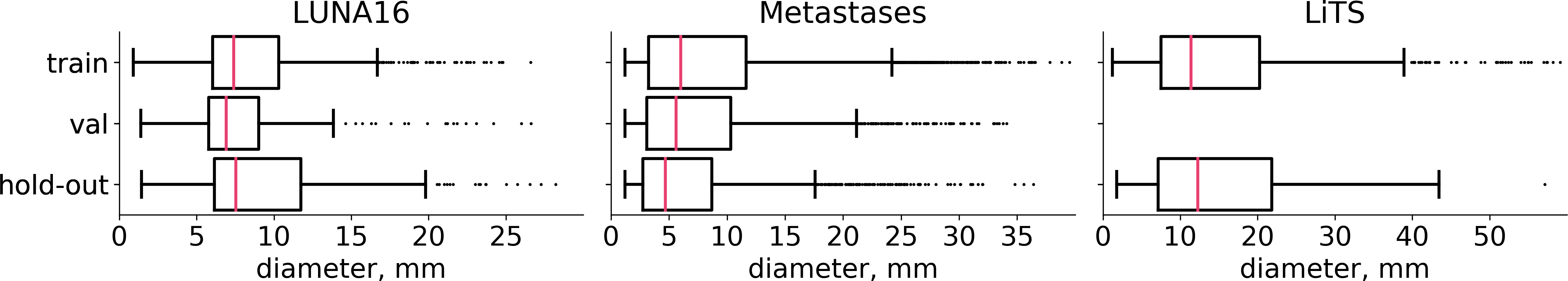}
    \caption{Lesion diameters distribution. Lung nodules under $10$ mm, metastases under $5$ mm and liver tumors under $12$ mm are considered \textit{small}, according to the clinical recommendations \cite{bankier2017recommendations,lin2015response}.}
    \label{fig:diameters}
\end{figure}

Learning a model under the presence of extremely small targets is challenging. This is especially the problem for 3D medical image segmentation tasks. The total fraction of voxels with lesion is about $0.1\%$ in the case of lung nodules and about $1\%$ in case of multiple brain metastases.
Moreover, in a series of medical image segmentation tasks we have a problem with the size imbalance. In some cases, large lesions could be up to 50 times bigger than the small ones (see typical lesion diameters distribution on Fig. \ref{fig:diameters}).

\begin{figure}[t]
    \begin{minipage}{0.32\linewidth}
        \centering
        BCE \\
        \includegraphics[width=\linewidth]{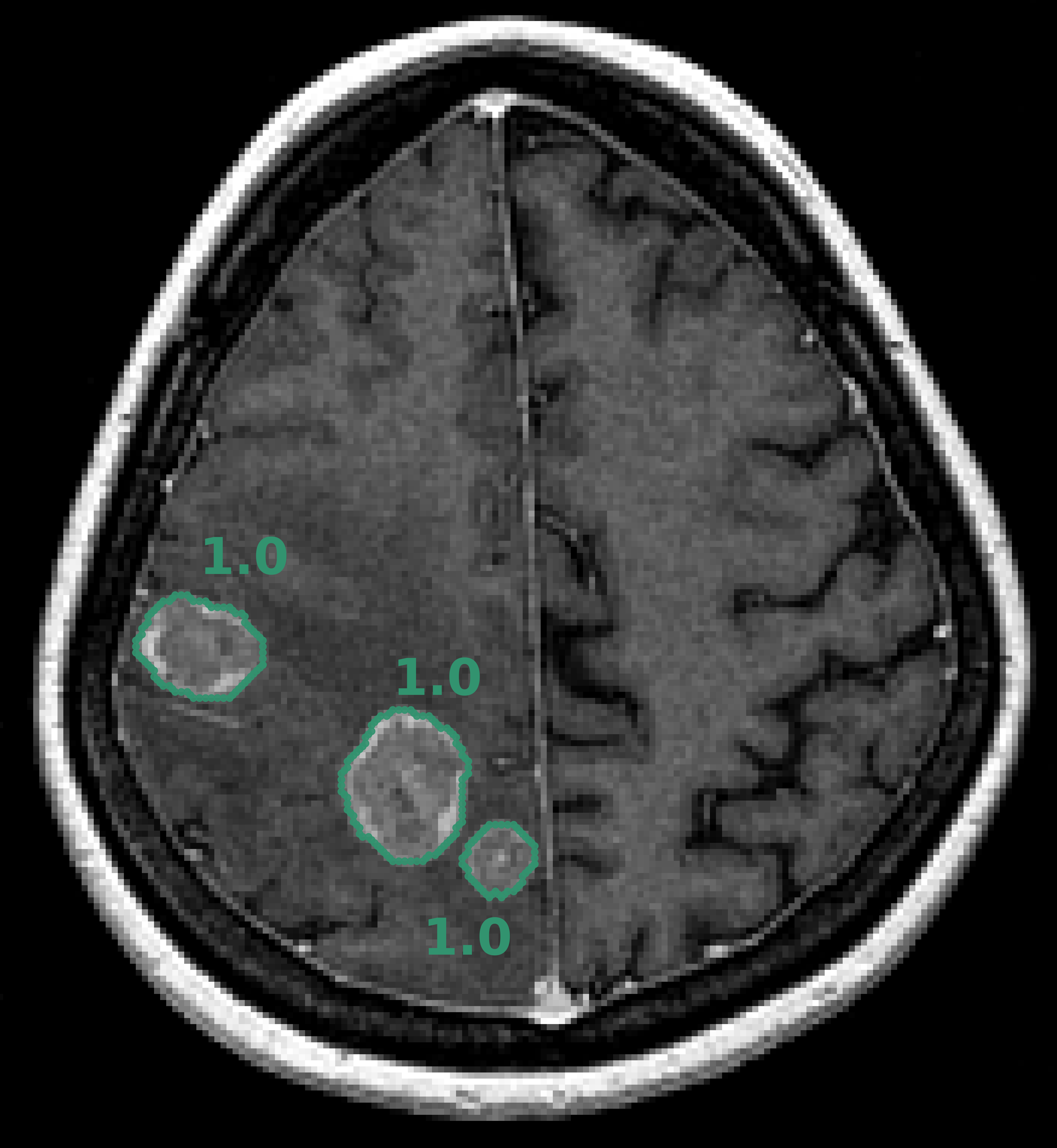} \\ \small Affected by imbalance
    \end{minipage}
    \hfill
    \begin{minipage}{0.32\linewidth}
        \centering
        WCE \\
        \includegraphics[width=\linewidth]{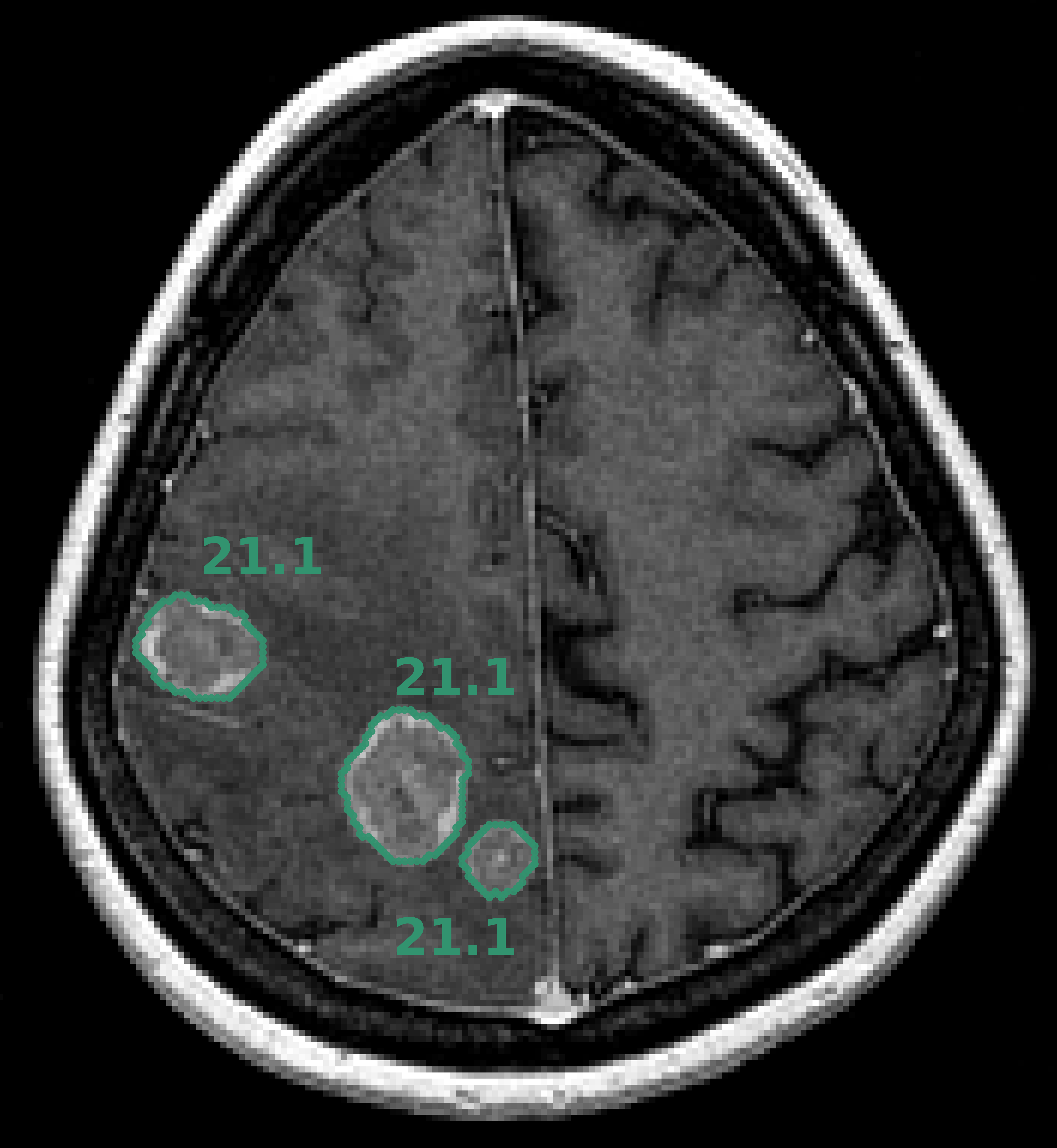} \\ \small Solving \textbf{class} imbalance
    \end{minipage}
    \hfill
    \begin{minipage}{0.32\linewidth}
        \centering
        \textbf{iw}BCE
        \includegraphics[width=\linewidth]{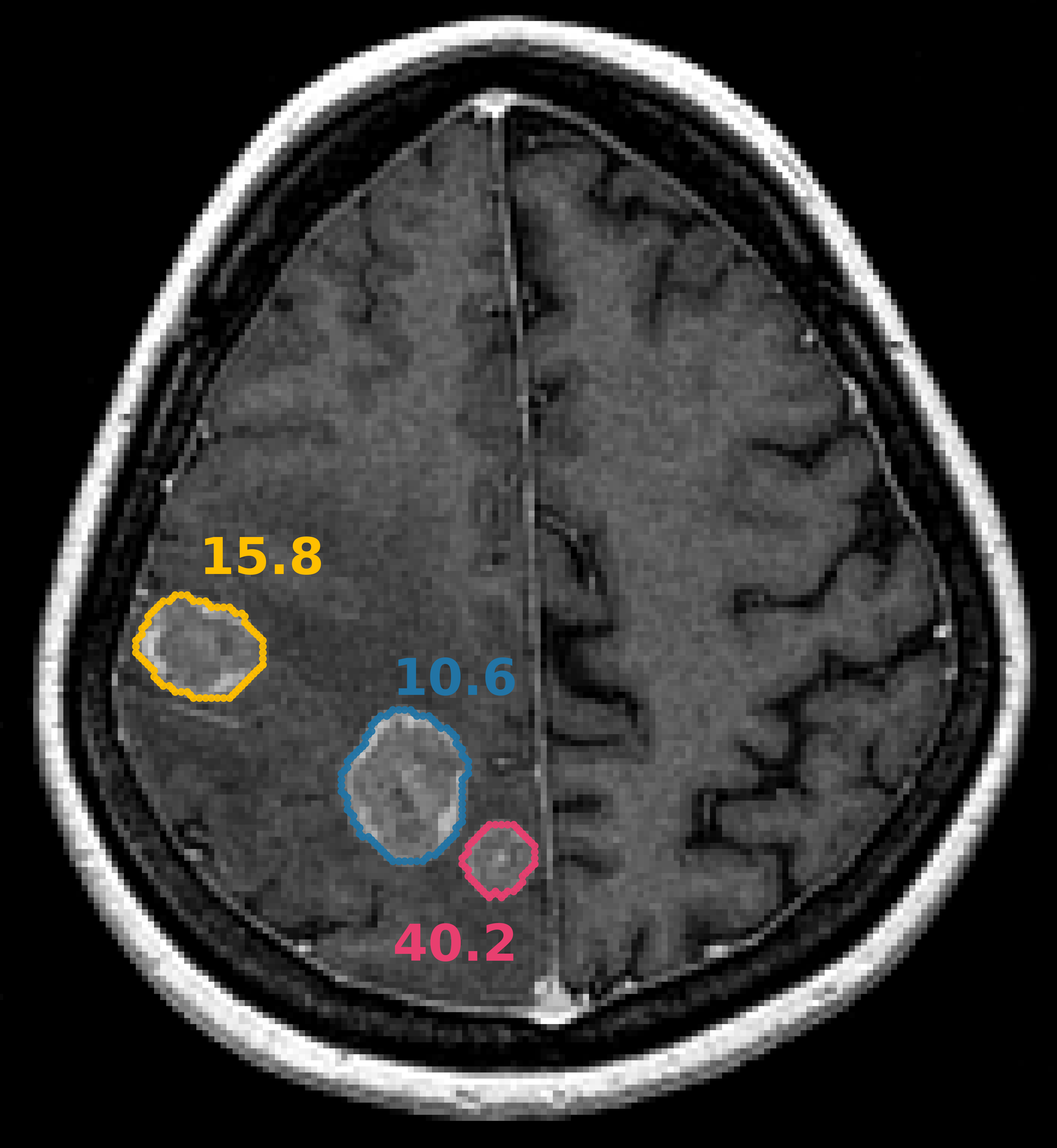} \\ \small Solving \textbf{size} imbalance
    \end{minipage}
    \caption{The effect of inverse weighting. No reweighting applied (left), class balancing via Weighted-Cross Entropy (center), \textbf{inverse weighting} (right). Weights for every tumor are calculated using formulas in Tab. \ref{tab:loss_table} and placed near the tumors.}
    \label{fig:bce_wce_iwbce}
\end{figure}

Several approaches have been suggested to tackle the problem of target imbalance. The main idea is to add weight to a loss function to equally represent each class (lesion vs non-lesion or different lesion types in a multi-class problem). It is implemented, for example, in Weighted Cross-Entropy \cite{unet} and Generalized Dice Loss \cite{sudre2017generalised}. The shortcoming of this approach is that it pays attention only to the lesion type, but not the lesion size (see Fig. \ref{fig:bce_wce_iwbce}). Besides, most of the research focuses on the delineation quality and lacks an investigation into the detection performance. Ideal segmentation implies perfect detection, however, due to the substantial differences between large and small lesions, almost a perfect delineation could have poor detection quality. Here we address this problem by applying the idea of weighting a loss function with respect to target sizes.\\
\\
\let\labelitemi\labelitemii
\noindent Our contribution is twofold: 
\begin{itemize}
    \item We propose a loss function reweighting strategy, that balances the lesions of different sizes. We call our approach \textbf{inverse weighting}, since the generated weights are inversely proportional to the lesion size. 
    \item We evaluate the effect of using the most popular segmentation loss functions on segmentation quality and network’s ability to detect lesions of different sizes. On a series of medical image segmentation tasks, we show how our approach improves the detection quality, especially for small lesions (Fig. \ref{fig:delta}), while preserving delineation performance.
\end{itemize}


\section{Related work}
\label{sec:related_work}

A large number of neural network architectures, improved training procedures, and loss functions have been proposed in recent years. We extensively investigate the behavior of loss functions keeping the rest of the deep learning pipeline on the state-of-the-art level without diving into details. 

The \textit{Binary Cross-Entropy} (BCE) is the standard loss function commonly used for segmentation tasks. It does not handle the problem of class imbalance and differently sized objects thus often yielding poor results. Authors of \cite{lin2017focal} suggested using \textit{Focal Loss} as an extension of BCE in highly class imbalanced detection tasks and it is widely used in segmentation tasks as well \cite{hashemi2018asymmetric}. Focal Loss does not apply any type of reweighting but automatically focuses the network attention on difficult examples. \textit{Dice Loss} \cite{milletari2016v} has recently become one of the state-of-the-art losses for medical image segmentation tasks. The authors claim that Dice Loss establishes the right balance between classes without assigning any weights. But for the tasks with multiple targets, a large object overshadows the small one, hence the network tends to miss small lesions. Recent work \cite{hashemi2018asymmetric} proposed \textit{Asymmetric Similarity Loss} (ASL) based on $F_{\beta}$ score. ASL extends Dice Loss (the special case with $\beta = 1$) and allows training a network with a better balance between precision and recall. But it shares the same drawback with Dice Loss: differently sized overshadowing objects. Authors of \cite{brosch2015deep} proposed Sensitivity-Specificity loss which we left without consideration. It performs worse than Dice Loss on a 3D medical image segmentation task in \cite{sudre2017generalised} and utilizes a similar idea with ASL.

Several approaches reweight BCE and Dice Loss to improve network performance in medical image segmentation tasks. In \cite{unet} authors use \textit{Weighted Cross-Entropy} (WCE) loss and \cite{sudre2017generalised} suggest Generalized Dice Loss (GDL) to tackle the problem of class imbalance. Both approaches utilize the same idea of reweighting the corresponding losses with weights inverse to the sizes of classes (see Tab. \ref{tab:loss_table}). Our approach simultaneously solves class imbalance problem and imbalance between differently sized objects. A deeper modification of Cross-Entropy loss to handle class imbalance is evaluated in \cite{li2019overfitting}, but the goal is quite different -- overfitting on small datasets. In \cite{wong20183d} authors suggest, a highly dependable on hyperparameters, a combination of Cross-Entropy and logarithmic Dice Loss to solve multiclass (19 classes) segmentation problem. In our work, we show an improvement for both of these losses independently.

We focus our attention on the most relevant loss functions and their explicitly reweighted modifications. Below we detail how our method is applied to state-of-the-art losses and compare it with WCE and GDL.

\section{Method}
\label{sec:method}

We find out that all models tend to miss small targets when training with BCE or Focal Loss. We assume poor performance comes from the inability of these losses to equally represent differently sized targets. Dice Loss and ASL have the same drawback: large targets overshadow the small ones. Moreover, already developed losses handle only the imbalance between classes, not between lesion sizes. We aim to close the gap and propose a simple methodology to reweight loss functions in the way that all targets contribute equally, e.g. small targets have greater weights.

During the training stage, we generate a tensor of weights for every incoming patch. To form such a tensor we split the corresponding ground-truth patch into $K+1$ connected components $L_0,\ldots, L_K$, where $L_0$ is the non-lesion component (background) and $K$ is the number of lesions in the current patch. Next, we assign the weight to every component which is inverse to the component's volume:

\begin{equation}
\label{eq:iw}
    w_j = \dfrac{\sum_{k=0}^K |L_k|}{\left( K + 1 \right) \cdot |L_j|},
\end{equation}
here $w_j$ is the weight, assigned to every voxel inside the corresponding component $L_j$. The constant in the denominator ensures that the sum of our weights is equal to the sum of the unit tensor of the same size (see derivation details in Supplementary Materials).
We call this method \textbf{inverse weighting (iw)}. Note,  how  our approach  assigns  greater  weights  to  the smaller tumors (Fig. \ref{fig:bce_wce_iwbce}). 
At this point, we can modify any of the discussed loss functions with our reweighting. Since WCE and GDL explicitly reweight state-of-the-art losses, we do not apply reweighting twice. Corresponding modifications for BCE, Focal Loss, Dice Loss, and ASL are shown in the Tab. \ref{tab:loss_table}.
    

\begin{table}
    \centering
    \caption{Loss functions and their modifications. Here $y_i$ denotes the $i^{th}$ element of the ground truth binary mask, $p_i$ is the corresponding predicted probability, and $\mathbf{w_i}$ is the proposed inverse weight. 
    }
    \begin{tabular}{l C{5cm} C{5cm}}
    \toprule
        Loss & Original Expression & Proposed Modification (\textbf{iw}) \\
        \midrule
        
        \scriptsize BCE & \scriptsize \(\displaystyle -(y_i\!\log p_i\!+\!(1\!-\!y_i)\!\log (1\!-\!p_i))\) & \scriptsize \( \displaystyle -\mathbf{w_i}(y_i\!\log p_i\!+\!(1\!-\!y_i)\!\log (1\!-\!p_i))\) \\
        
        \cmidrule{2-3}
        
        \scriptsize \multirow{2}{*}{$\text{Focal Loss}_{\gamma, \alpha}$} & \multicolumn{1}{l}{\scriptsize \(\displaystyle -\!(\alpha(1\!-\!p_i)^{\gamma} y_i\!\log\!p_i \)} & \multicolumn{1}{l}{\scriptsize \(\displaystyle \mathbf{w_i}(\alpha(1\!-\!p_i)^{\gamma} y_i\!\log p_i \)}\\
        
        & \multicolumn{1}{r}{\scriptsize \(\displaystyle + (1\!-\!\alpha)p_i^{\gamma} (1\!-\!y_i\!)\!\log (1\!-\!p_i))\)} & \multicolumn{1}{r}{\scriptsize \(\displaystyle + (1\!-\!\alpha)p_i^{\gamma} (1\!-\!y_i)\!\log (1\!-\!p_i))\)}\\
        
        \cmidrule{2-3}
        
        \scriptsize WCE & \scriptsize \(\displaystyle -w y_i\!\log p_i-(1\!-\!y_i\!)\!\log (1\!-\!p_i\!), \smallskip w\!=\!\!\frac{n\!\!-\!\!\sum_i\!p_i}{\sum_i\!p_i} \) & ---\\
        
        \midrule
        
        \scriptsize Dice Loss & \scriptsize \(\displaystyle 1\!-\!\frac{2 \sum_i\!p_{i}\!y_{i}}{\sum_i\!(p_{i}^2\!+\!y_{i}^2)} \) & \scriptsize \(\displaystyle  1\!-\!\frac{2\sum_i\!\mathbf{w_i}p_i\!y_i}{ \sum_i\!\mathbf{w_i}(p_i^2\!+\!y_i^2)} \) \\
        
        \cmidrule{2-3}
        
        \scriptsize $\text{ASL}_{\beta}$ & \scriptsize \(\displaystyle 1\!-\!\frac{(1\!+\!\beta^2)\!\sum_i\!p_i\!y_i}{\sum_i\!(\beta^2\!y_i\!+\!p_i)} \)
        & \scriptsize \(\displaystyle 1\!-\!\frac{(1\!+\!\beta^2)\!\sum_i\!\mathbf{w_i}p_i\!y_i}{\sum_i\!\mathbf{w_i}(\beta^2\!y_i\!+\!p_i)} \) \\
        
        \cmidrule{2-3}
        
        \scriptsize GDL & \scriptsize \(\displaystyle  1\!-\!\frac{2\sum_{c=1}^2\!w_c^2\!\sum_{i}\!p_{i} y_{i}}{\sum_{c=1}^2\!w_c^2\!\sum_{i}\!(p^2_{i}\!+\!y^2_{i})}, \smallskip w_c\!=\!\!\frac{1}{\sum_{i}\!y_{i}} \) & --- \\
        
        \bottomrule
    \end{tabular}
    
    \label{tab:loss_table}
\end{table}

\section{Experiments}
\label{sec:exp}

\subsection{Data}
\label{ssec:data}

We report our results on three datasets. Two publicly available datasets that include 3D CT images: LUNA16 \cite{luna16} with lung cancerous nodules and LiTS \cite{bilic2019liver} with liver tumors; and one private dataset with MR images of multiple brain metastases.

\textbf{LUNA16} includes $816$ (we have excluded $72$ cases with nodules located outside of lung masks) annotated chest scans from LIDC/IDRI database \cite{armato2011lung}. For every image, we clip intensities between $-1000$ and $300$ Hounsfield units (HU), and then set the voxels outside the given binary lung masks to $-1000$. Ground truth mask was formed by averaging 4 given annotations.

\textbf{Metastases} (private dataset) includes $1952$ unique patients with the T1-weighted MRI of the head. We apply no preprocessing steps to these images.

\textbf{LiTS} includes $131$ annotated CT abdomen scans. For every image, we clip intensities between $-300$ and $300$ HU and then apply a given binary mask of liver the same way we did it with LUNA16 data.

Before passing through the network, we scale images to have voxel's intensities between $0$ and $1$.





We use \textit{train}-\textit{validation} setup to compare different architectures and hyperparameters for loss functions. Then the merged combination of \textit{training} and \textit{validation} data is used to train the chosen methods and we report final results on previously unseen \textit{hold-out} set. LUNA16 is presented as $10$, approximately equal, subsets \cite{luna16} thus we use the first $6$ for \textit{training} ($534$ images), next $2$ for \textit{validation} ($178$ images) and the last pair as \textit{hold-out} ($174$ images). We divide Metastases into \textit{training} ($1250$ images), \textit{validation} ($402$ images) and \textit{hold-out} ($300$ images). LiTS is also presented as $2$ subsets, so we use the first for \textit{training} ($104$ images) and the second as \textit{hold-out} ($27$ images). We do not shrink the validation part of the LiTS, since this dataset is used only once for the final results reporting.

\subsection{Architecture and training}
\label{ssec:setup}

For all our experiments we consistently use a single CNN model -- slightly modified 3D U-Net \cite{unet3d}. Implemented architecture within PyTorch framework is available in our repository along with a schematic image. Following the suggestion of \cite{isensee2018no}, we do not focus our attention on fine-tuning the CNN model.


In all scenarios we train the model for $100$ epochs, starting with learning rate of $10^{-2}$, and reducing it to $10^{-3}$ at the epoch $80$. Each epoch consists of $100$ iterations of stochastic gradient descent with Nesterov momentum ($0.9$). At every iteration we sample patches of size $128 \times 128 \times 128$ and batch size of two. With the probability of $0.5$ we sample the patch so that it contains at least one voxel with lesion, otherwise we sample it uniformly. The training takes about $26$ hours on a 24GB NVIDIA Tesla M40 GPU.


Note, that only two of the considered loss functions have hyperparameters: ASL ($\beta$) and Focal Loss ($\gamma, \alpha$). We use ASL with $\beta = 1.5$ originally recommended in \cite{hashemi2018asymmetric}. For Focal Loss we also use $\gamma = 2$ originally recommended in \cite{lin2017focal}, but change $\alpha$ to be $0.75$ chosen on validation. 

\subsection{Metric}
\label{ssec:metric}

Dice Score has a particular drawback measuring the delineation quality in the tasks with multiple lesions per image: big lesion overshadows small ones. We use \textbf{object Dice Score} -- the average Dice Score over \textit{unique found lesions}. Therefore it does not shift towards larger lesions. Note that we exclude missed lesions from this analysis, hence the delineation quality is independent from detection quality.


To measure the detection quality we suggest using a Free-response Receiver Operating Characteristic (FROC) curve analysis. It is extremely efficient operating with multiple targets and False Positive (FP) responses per case \cite{deluca2008extensions}. A FROC curve measures the sensitivity to detected objects instead of voxel-wise sensitivity, therefore does not have the same drawback of overshadowed lesions. A FROC curve summarizes the model's efficiency with the trade-off between the fraction of lesions detected (Recall) and the average number of FPs per image. But it gives us only visual representation of experimental results. To compare the performance of different methods we extract a single value from the curves. Authors of \cite{van2010comparing} suggested using the \textbf{average object-wise Recall} over the predefined FP values ($1/8$, $1/4$, $1/2$, $1$, $2$, $4$, $8$) which is also the main metric of LUNA16 challenge \cite{luna16}. This metric gives us the average fraction of detected lesions per case which is highly interpretable in terms of detection quality. 

To calculate the confidence intervals for FROC curves and for average Recall we use bootstrapping. We sample 80\% of test patients and build a curve on every of the 100 iterations. Average recall is calculated for every bootstrapped curve and we report the mean value along with the standard deviation.

\subsection{Results and discussion}
\label{ssec:results}

\begin{figure}[h]
      \includegraphics[width=\linewidth]{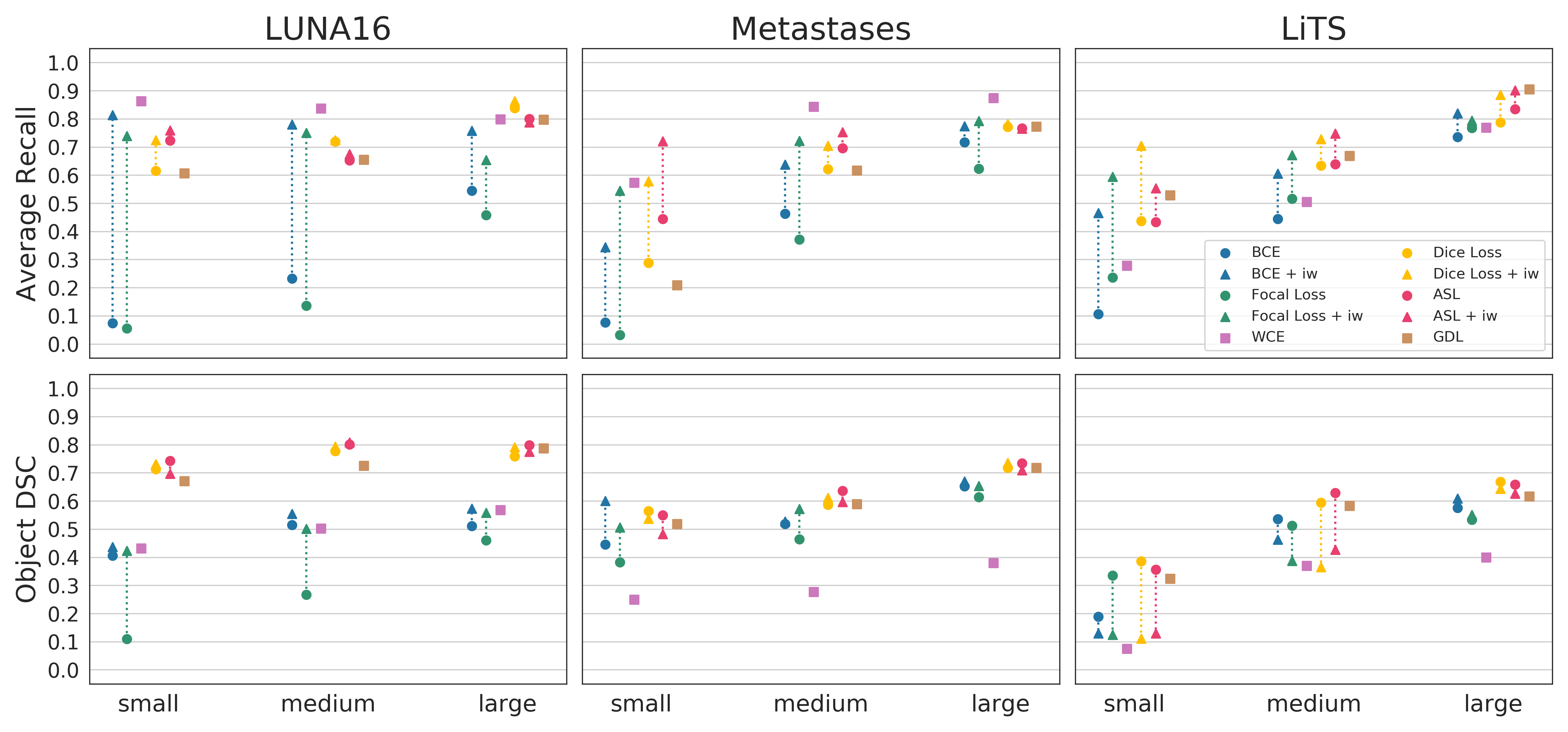}
      \caption{The impact of inversely weighted loss functions in terms of average Recall and object-wise Dice Score. We show performance on three approximately equal subsets ($1/3$ each) of lesions divided by their size. Small and medium groups correspond to the clinical recommendations of small lesions (see Fig. \ref{fig:diameters}).}
      \label{fig:delta}
\end{figure}

We visualize our main contribution with the considerable improvement of the average object-wise Recall for all four chosen loss functions on all three datasets (Fig. \ref{fig:delta}). We also report our metrics separately for three groups of lesion sizes and show a solid contribution into the small lesion detection quality which satisfies our method's motivation. However, a comparison with WCE is worth a more detailed discussion.


Images from LUNA16 contain $1.3$ nodules per scan on average, while Metastases and LiTS have $4.8$ and $6.9$ tumors per scan respectively. The latter means that LUNA16 is hardly an appropriate dataset to benefit from our method, since the majority of training patches contain only one lesion. One lesion per patch is clearly the \textit{class imbalance} problem, and WCE outperforms the other methods in terms of average Recall. But nevertheless we show inverse weighting solving also the class imbalance task on the competitive level solidly improving BCE and Focal Loss performance. Finally, even the slight improvement in the detection quality of WCE comes with the dramatic delineation quality loss on the other two datasets, which is crucial for clinical tasks.

GDL failed to surpass inversely weighted loss function almost in all scenarios. But overall we find ASL and Dice Loss along with GDL and their inversely weighted modifications to be highly stable during the training. Respectively, Dice-like loss function sufficiently outperform BCE-like losses both in terms of the detection and the delineation qualities. We believe such a behaviour comes from two properties of Dice Loss. Firstly, it is designed to optimize the Dice Score metric, and one could clearly see the dominance of Dice-like losses in terms of object Dice Score (Fig. \ref{fig:delta} and Tab. \ref{tab:res}). Secondly, it partially solves the class imbalance problem, but only in the cases with exactly one object per patch. The latter is again perfectly demonstrated on LUNA16, as we put this dataset to be more about class imbalance problem in the previous paragraph. One could see the already high object Dice Scores and average Recall values of ASL and Dice Loss on LUNA16 along with minor changes of their reweighting.

However, modified with inverse weighting loss functions have a noticeable decrease in delineation quality on LiTS data. We consider this to be a side effect of highly increased object-wise Recall: \textit{modified losses find more difficult cases, hence joint object Dice Score could decrease}.

Besides the separate performance on lesion sizes we also include more detailed results for all lesions in hold-out sets (Tab. \ref{tab:res}).
We give the visual representation of experimental results in terms of detection quality via FROC analysis (see Supplementary Materials, Fig. \ref{fig:froc}).

\begin{table}[h]
    \centering
    \caption{Results for all considered loss functions along with the proposed method -- inverse weighting ("$+$" with iw, "$-$" without iw). The numbers in brackets are standard deviation. 
    }
    \begin{tabular}{l c C{1.59cm} C{1.59cm} C{1.59cm} C{1.59cm} C{1.59cm} C{1.59cm}}
        \toprule
        & \multirow{2}{*}{\textbf{iw}} & \multicolumn{2}{c}{LUNA16} & \multicolumn{2}{c}{Metastases} &
        \multicolumn{2}{c}{LiTS}\\
        \cmidrule(lr){3-4} \cmidrule(lr){5-6} \cmidrule(lr){7-8}
        &  & avg Recall & obj DSC & avg Recall & obj DSC & avg Recall & obj DSC \\
        
        \midrule
        \multirow{2}{*}{BCE} 
        & $-$ & .42 (.02) & .57 (.28) & .47 (.01) & .67 (.25) & .47 (.03) & .61 (.29) \\
        & $+$ & .67 (.01) & .56 (.20) & .52 (.01) & .66 (.23) & .59 (.03) & .53 (.27) \\
                             
        \cmidrule{2-8}
        \multirow{2}{*}{Focal Loss} 
        & $-$ & .35 (.02) & .51 (.28) & .40 (.01) & .64 (.25) & .50 (.03) & .58 (.27) \\
        & $+$ & .55 (.01) & .54 (.20) & .52 (.01) & .63 (.21) & .62 (.03) & .48 (.27) \\
        
        \cmidrule{2-8}
        WCE & $-$ & .74 (.01) & .50 (.17) & .54 (.01) & .39 (.22) & .52 (.04) & .41 (.29) \\
        
        \midrule
        
        \multirow{2}{*}{Dice Loss} 
        & $-$ & .71 (.02) & .76 (.20) & .55 (.01) & .69 (.23) & .62 (.03) & .63 (.25) \\
        & $+$ & .73 (.02) & .77 (.16) & .57 (.01) & .68 (.21) & .72 (.03) & .49 (.30) \\
        
        \cmidrule{2-8}
        \multirow{2}{*}{ASL}
        & $-$ & .68 (.02) & .77 (.16) & .55 (.01) & .71 (.20) & .66 (.03) & .63 (.24) \\
        & $+$ & .70 (.02) & .76 (.18) & .59 (.02) & .66 (.20) & .73 (.03) & .53 (.28) \\
        
        \cmidrule{2-8}
        GDL & $-$ & .69 (.02) & .73 (.20) & .53 (.01) & .70 (.22) & .69 (.03) & .60 (.27) \\
        
        \bottomrule
    \end{tabular}
    \label{tab:res}
\end{table}

\section{Conclusion}
\label{sec:conclusion}

We propose a universal approach to loss functions reweighting. It could be used with almost any state-of-the-art loss function. Our experiment demonstrates an improvement of network's ability to detect lesions for Cross-Entropy, Focal Loss, Dice Loss and Asymmetric Similarity Loss on three medical tasks with multiple targets per case. Moreover, we believe the method can also improve quality with other complex multi-stage pipelines or with any other CNN architecture which is the goal for our future research.



\paragraph{Acknowledgements.}


The results of the paper are based on the scientific research supported by the Russian Science Foundation under grant 17-11-01390. The authors also acknowledge the National Cancer Institute and the Foundation for the National Institutes of Health, and their critical role in the creation of the free publicly available LIDC/IDRI Database used in this study.

\bibliographystyle{splncs04}
\bibliography{bibliography1600.bib}

\section*{Supplementary materials}

\subsection*{Free-response Receiver Operating Characteristic curve analysis}
\label{ssec:img}

\begin{figure}[H]
\centering
\includegraphics[width=\linewidth]{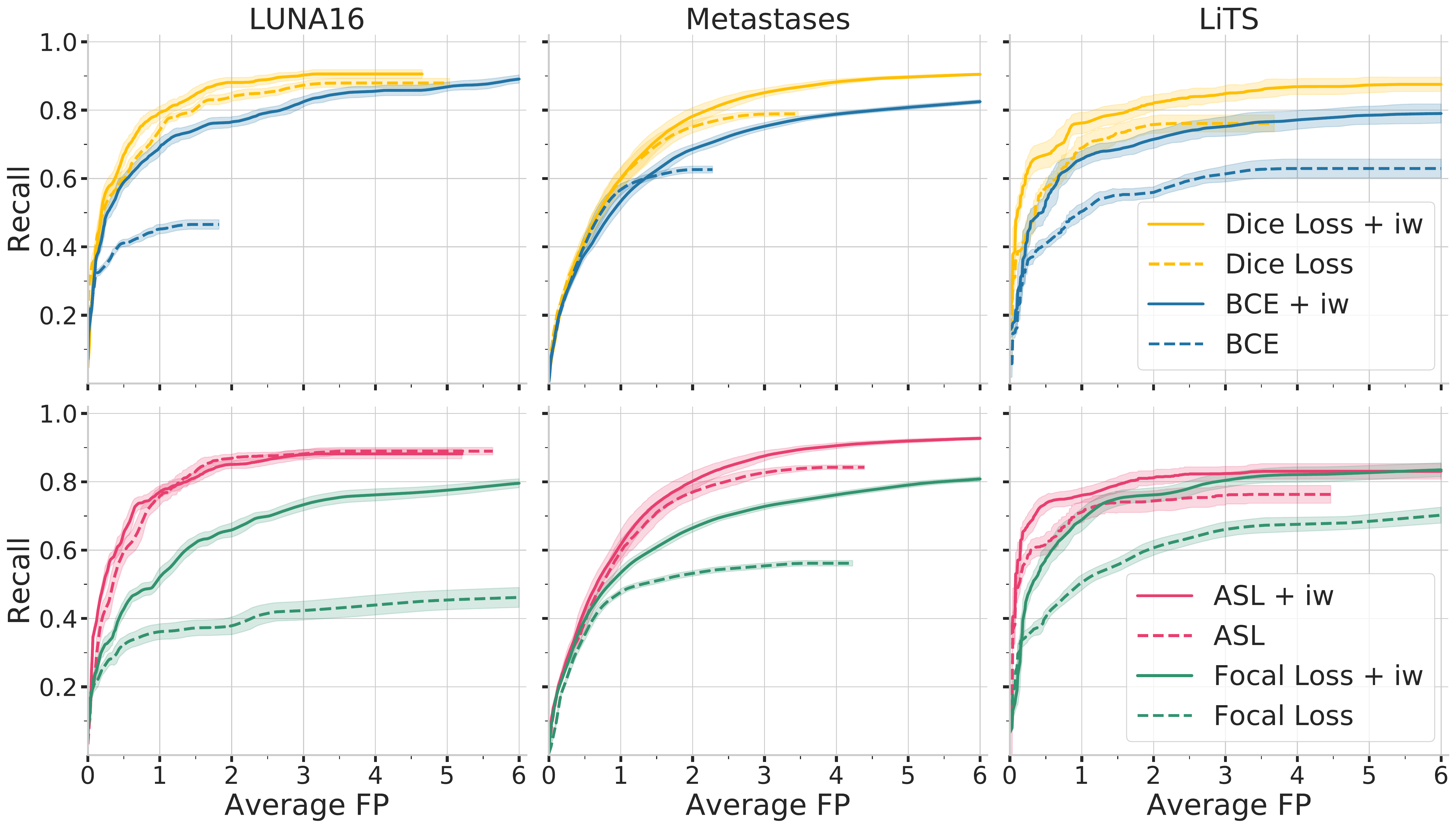}
\caption{The impact of the proposal in terms of FROC curve analysis for all three utilized datasets and all lesion sizes jointly. We show an improvement for every loss function (dashed lines) with the proposed \textbf{inverse weighting} (solid lines). The shadowed area corresponds to the standard deviation along the Y-axis.}
\label{fig:froc}
\end{figure}

\subsection*{Inverse weighting derivation}
\label{ssec:rew}
\textbf{Goal}: given $K$ separate lesions $L_1,\ldots, L_K$ on the image and a non-lesion component (background) $L_0$, we want them approximately equally contribute to a loss function (e.g.  Binary Cross-Entropy):

\begin{equation}
\label{eq:1}
    \sum_{i \in L_k} - w_k \log p_i = \sum_{j \in L_m} - w_m \log p_j, \; \forall k, m \in \{0,\ldots, K\},
\end{equation}
\noindent here $w_k$ is the weight assigned to every voxel of $L_k$ (every separate lesion gets its own weight), and $p_i$ is the estimated probability of the corresponding voxel class.
\underline{Assuming constant prediction}, i.e. all probabilities are equal to $p$, Eq. \ref{eq:1} becomes:

\begin{equation}
\label{eq:2}
    | L_k |  w_k = | L_m | w_m , \; \forall k, m \in \{0,\ldots, K\}.
\end{equation}
Now, after adding a normalization condition $\sum_{n=1}^N w_n = N$, where N is the number of voxels inside the patch, we can derive $w_0$:

\begin{equation} \label{eq3}
\begin{split}
    N & =\sum_{n=1}^N w_n\\
      & = \sum_{k=0}^K \smallskip \sum_{i \in L_k} w_i\\
      & = \sum_{k=0}^K | L_k | w_k \\
      & = \sum_{k=0}^K |L_0| w_0 \\
      & = \left( K + 1 \right) |L_0| w_0,
\end{split}
\end{equation}
therefore:
\begin{equation}
\label{eq:3}
    w_0 = \frac{N}{\left( K + 1 \right) \cdot |L_0|}.
\end{equation}

\noindent Now, by combining Eq. \ref{eq:2} and Eq. \ref{eq:3} we finally get:

\begin{equation} \label{eq:4}
\begin{split}
    w_j &  = \frac{N}{\left( K + 1 \right) \cdot |L_j|} \\ 
    & = \frac{\sum_{k=0}^K |L_k|}{\left( K + 1 \right) \cdot |L_j|}.
\end{split}
\end{equation}

\end{document}